\documentclass[conference]{IEEEtran}
\IEEEoverridecommandlockouts
\usepackage[nocompress]{cite}
\usepackage{amsmath,amssymb,amsfonts}
\usepackage{algorithmic}
\usepackage{graphicx}
\usepackage{textcomp}
\usepackage{xcolor}
\usepackage{siunitx}
\DeclareSIUnit\torr{Torr}

\begin{document}

\title{Multi-beam RF accelerators for ion implantation
\thanks{This work was supported by the US Department of Energy through the
ARPA-E ALPHA program and the Energy I-Corps program under contract DE-AC0205CH11231.}}

\author{\IEEEauthorblockN{Peter A. Seidl}
\IEEEauthorblockA{\textit{Lawrence Berkeley National Laboratory}\\
Berkeley, CA 94720, USA \\
paseidl@lbl.gov}
\and
\IEEEauthorblockN{Arun Persaud}
\IEEEauthorblockA{\textit{Lawrence Berkeley National Laboratory}\\
Berkeley, CA 94720, USA \\
apersaud@lbl.gov}
\and
\IEEEauthorblockN{Diego Di Domenico}
\IEEEauthorblockA{\textit{School of Engineering and Architecture} \\
Fribourg, Switzerland \\
diego.didomenico@outlook.com}
\and
\IEEEauthorblockN{Johan Andreasson}
\IEEEauthorblockA{\textit{Airity Technologies, Inc.} \\
Redwood City, CA, 94061 USA \\
johan@airitytech.com}
\and
\IEEEauthorblockN{Qing Ji}
\IEEEauthorblockA{
\textit{Lawrence Berkeley National Laboratory}\\
Berkeley, CA 94720, USA \\
qji@lbl.gov}
\and
\IEEEauthorblockN{Wei Liang}
\IEEEauthorblockA{\textit{Airity Technologies, Inc.} \\
Redwood City, CA, 94061 USA \\
wei@airitytech.com}
\and
\IEEEauthorblockN{Di Ni}
\IEEEauthorblockA{\textit{SonicMEMS Laboratory} \\
\textit{Cornell University}\\
Ithaca, NY 14853, USA \\
dn273@cornell.edu}
\and
\IEEEauthorblockN{Daniel Oberson}
\IEEEauthorblockA{\textit{School of Engineering and Architecture of Fribourg} \\
Fribourg, Switzerland \\
daniel.oberson@hefr.ch}
\and
\IEEEauthorblockN{Luke Raymond}
\IEEEauthorblockA{\textit{Airity Technologies, Inc.} \\
Redwood City, CA, 94061 USA \\
luke@airitytech.com}
\and
\IEEEauthorblockN{Gregory Scharfstein}
\IEEEauthorblockA{
\textit{Lawrence Berkeley National Laboratory}\\
Berkeley, CA 94720, USA \\
GAScharfstein@lbl.gov}
\and
\IEEEauthorblockN{Alan M.M. Todd}
\IEEEauthorblockA{\textit{AMMTodd Consulting} \\
Princeton Junction, NJ 08550 USA \\
AMMTodd@gmail.com}
\and
\IEEEauthorblockN{Amit Lal}
\IEEEauthorblockA{\textit{SonicMEMS Laboratory} \\
\textit{Cornell University}\\
Ithaca, NY 14853, USA \\
amit.lal@cornell.edu}
\and
\IEEEauthorblockN{Thomas Schenkel}
\IEEEauthorblockA{
\textit{Lawrence Berkeley National Laboratory}\\
Berkeley, CA 94720, USA \\
t\_schenkel@lbl.gov}
}

\maketitle

\begin{abstract}
We report on the development of a radio frequency (RF) linear accelerator (linac) for multiple-ion
beams that is made from stacks of low cost wafers. The accelerator lattice is comprised of
RF-acceleration gaps and electrostatic quadrupole focusing elements that are fabricated on
4" wafers made from printed circuit board or silicon. We demonstrate ion acceleration with
an effective gradient of about \SI{0.5}{\mega\volt\per\m} with an array of $\mathbf{3\times3}$
beams. The total ion beam energies achieved to date are in the \SI{10}{\keV} range with 
total ion currents in tests with noble gases of $\SI{\sim0.1}{\mA}$. We discuss scaling
of the ion energy (by adding acceleration modules) and ion currents (with more beams) for
applications of this multi-beam RF linac technology to ion implantation and surface 
modification of materials.  
\end{abstract}


\section{Introduction}
Ion implantation is a mature technology \cite{Hamm12}. In most implanters, a single ion beam is
formed by extraction of ions from an ion source. Ions are then accelerated and a desired species
is selected for implantation. The achievable ion current density is limited by space charge forces
and the total ion current is limited by the size of the extraction aperture from which ions
can be extracted to form a beam with low enough emittance for efficient transport in the beam line
\cite{Brown2004}. The concept of multi-beam ion accelerators was developed in the late 1970s by
Maschke et al. with the concept of a Multiple Electrostatic Quadrupole Linear Accelerator (MEQALAC)
\cite{Maschke_1979b}. MEQALACs are RF-driven linear accelerators where the total ion current can be
scaled by adding more beams and the ion kinetic energy can be increased by adding accelerator modules.
In the first implementations, MEQALACs used RF-cavities to achieve ion acceleration with high
voltages driven at frequencies in the \SI{25}{\MHz} range \cite{URBANUS1989508}.  Multi-beam 
RF-linacs can boost the total current for applications where beams of mass analyzed single species
are required. We have recently reported on the development of multi-beam RF accelerators that we
assemble from stacks of low cost wafers \cite{Persaud:RSI-2016,Seidl2018}. We form RF-acceleration
structures and electrostatic quadrupole (ESQ) focusing elements on printed circuit board and silicon
wafers with 10-cm diameter using standard microfabrication techniques \cite{Vinaya2018}. In this article
we report on the use of compact GaN based RF amplifiers to accelerate ions in an array of $3\times3$
beams with energy gains of up to \SI{5.1}{\kilo\volt} per RF-acceleration unit.  This method for
generating RF-acceleration voltages leads to an effective acceleration gradient of about
\SI{0.3}{\mega\volt\per\m} in our multi-beam RF linac.  While this is much lower than for 
conventional RF accelerators based on cavities, our implementation is very compact with high 
current density and is formed from low cost components in a modular architecture.  

\begin{figure}[htbp]
\centering
\includegraphics[width=\linewidth]{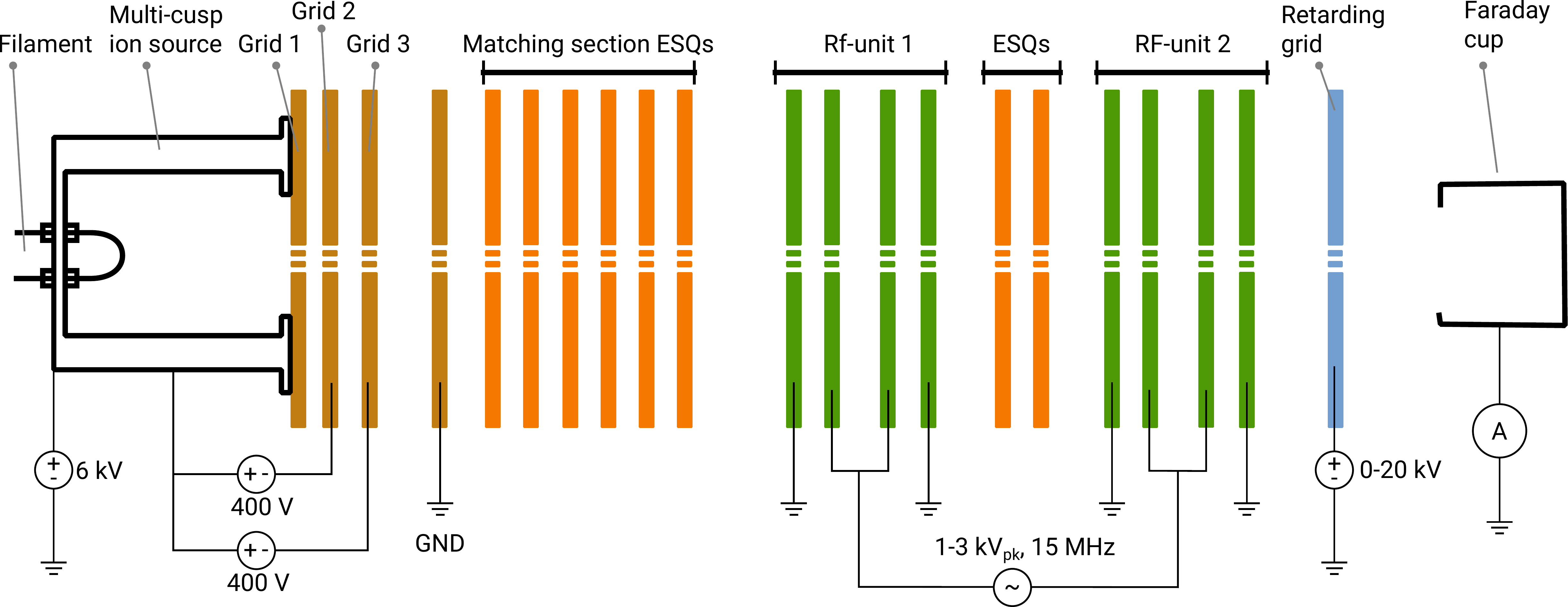}
\includegraphics[width=0.6\linewidth]{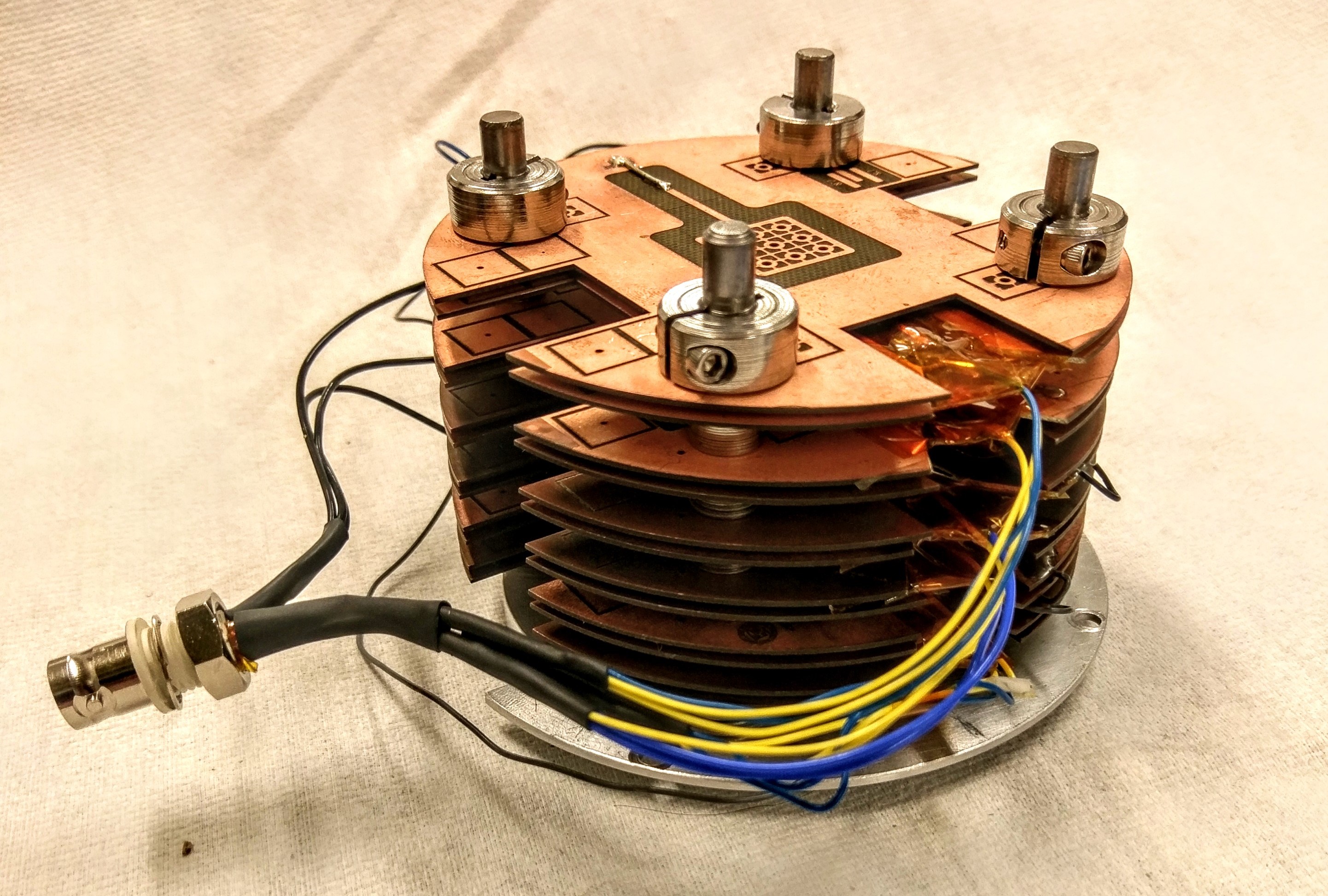}
\caption{Schematic (top) and photo (bottom) of the multi-beam RF linac with an array of
$3\times3$ beam apertures.}
\label{fig:concept}
\end{figure}

Figure~\ref{fig:concept} shows a schematic and a photo of the wafer based multi-beam RF-linac.
Ions are extracted from a $3\times3$ array of apertures (\SI{0.5}{\mm} diameter) from a filament driven
multi-cusp ion source and accelerated to 5 to \SI{10}{\keV}. We use argon and helium ions
for beam tests and the ion source can also provide ions of common dopant species. Following
a matching section comprised of a series of ESQ focusing elements, the ion beams reach the
lattice of RF acceleration units and ESQs. We determine the energy gain by scanning the
bias on a retarding grid while monitoring the ion current in a Faraday cup.  

The RF acceleration units are formed from four wafers, with the outside wafers held at ground and the 
central two wafers driven by high voltage pulses at the selected RF frequency. Ions entering an
RF acceleration unit are accelerated by the RF voltage and then drift in the field-free gap 
between the two central wafers before further acceleration  in the gap between the third wafer
and the grounded last wafer in the RF acceleration unit. The drift distance is set to match the 
phase advance of the RF at a given frequency according to $x=0.5\,\, v/f$, where $x$ is the gap distance,
$v$ the ion velocity and $f$ the RF frequency.  For operation at \SI{13.5}{\MHz} with argon ions injected
at \SI{10}{\keV} ($v_{Ar}=\SI{2.2e5}{\meter\per\second}$) the drift distances are \SI{8}{\mm}. After several stages of acceleration the RF frequency
can be increased to compensate for increasing ion velocities and to keep drift distances small.
In our first proof-of-concept of this approach of multi-beam ion acceleration we used an
external LC-tank circuit to generate RF acceleration voltages of 600 V/gap \cite{Persaud:RSI-2016}.
We have now implemented compact GaN based RF sources \cite{Airity} that are placed near the
acceleration wafers inside the vacuum chamber. The high voltage pulses are delivered through
low capacitance wires to the RF acceleration wafers. One RF source can drive ion acceleration
in several modules with up to \SI{2.6}{\kV} per acceleration gap.   

\section{RF source}
\begin{figure}[htbp]
\centering
\includegraphics[width=\linewidth]{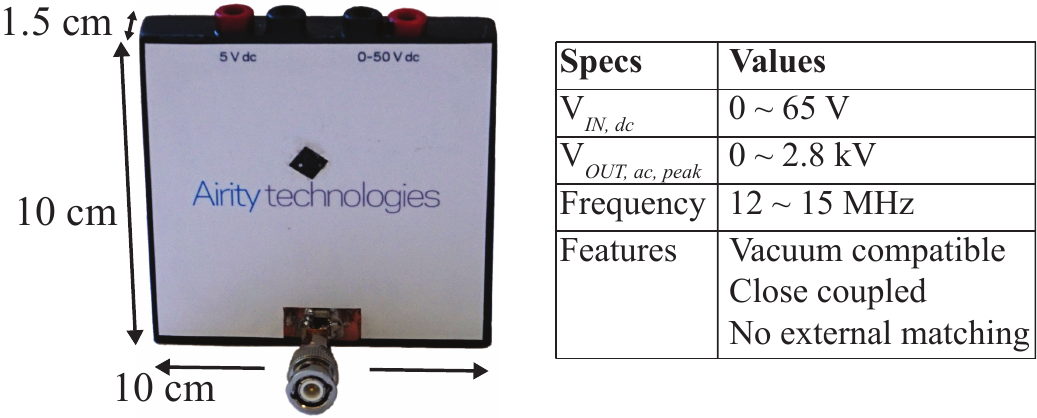}
\caption{Photo and specifications of the RF amplifier}
\label{fig:RFamplifier}
\end{figure}

A custom RF power source was developed by Airity technologies to generate the high
voltage AC required to drive the wafer array. By leveraging the intrinsic inductive and
capacitive elements of the wafers and incorporating them into the source design, a compact and efficient vacuum compatible RF source was built that enables placement close to the wafer array to minimize cable effects. The
specifications of the source are summarized in Figure ~\ref{fig:RFamplifier}.

Initial tests of an updated RF source design suggest that voltages exceeding
\SI{5}{\kV} are possible. Future work will further investigate such designs with
the goal of scaling array voltage to \SI{10}{\kV}.
\section{Modeling}
\begin{figure}[htbp]
\centering
\includegraphics[width=0.7\linewidth]{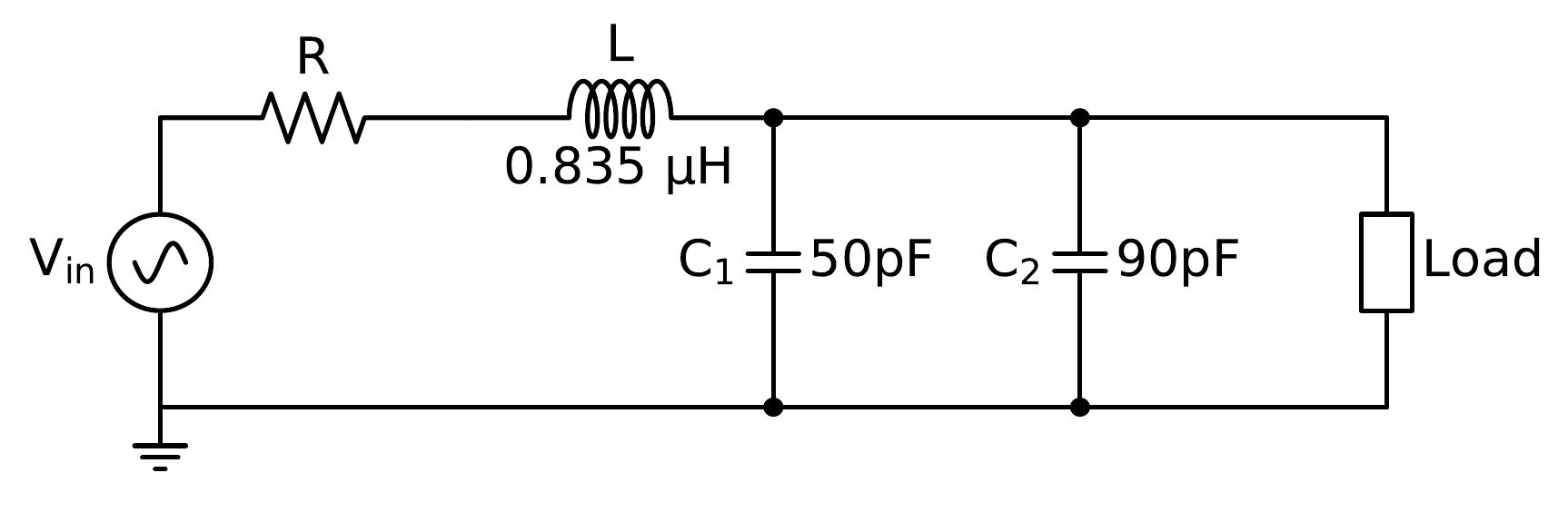}
\caption{The equivalent circuit used to model the RF source. The resistor
$R=(1.212\times\frac{V_{drive}}{\SI{103}{V}})\SI{}{\ohm}$ captures variations in regards to
the source voltage. The capacitor $C_1$ and inductor $L$ represent
internal values of the RF source. The load and $C_2$ represent the external load.}
\label{fig:source-schematic}
\end{figure}

To simulate the system and investigate its scaling behaviour an equivalent circuit was developed. 
The RF source was modeled as a simple $RLC$ circuit shown in Figure~\ref{fig:source-schematic}.
Since the RF source has been developed to drive a \SI{120}{\pico\farad} load, the capacitance
in the model has been split into an internal \SI{50}{\pico\farad} capacitance, $C_1$, and 
the external load. In our case our load of a single RF unit is smaller than the load for which the unit 
was designed and we therefore augmented the load capacitance with a parallel capacitor, $C_2$.
When driving more units this capacitor should be adjusted accordingly. The resistor $R$ in
the source model varies with the driving voltage. This was added to compensate for the observed
behaviour of lower gains at higher voltages, possibly due to heating of the RF driver.

\begin{figure}[htbp]
\centering
\includegraphics[width=\linewidth]{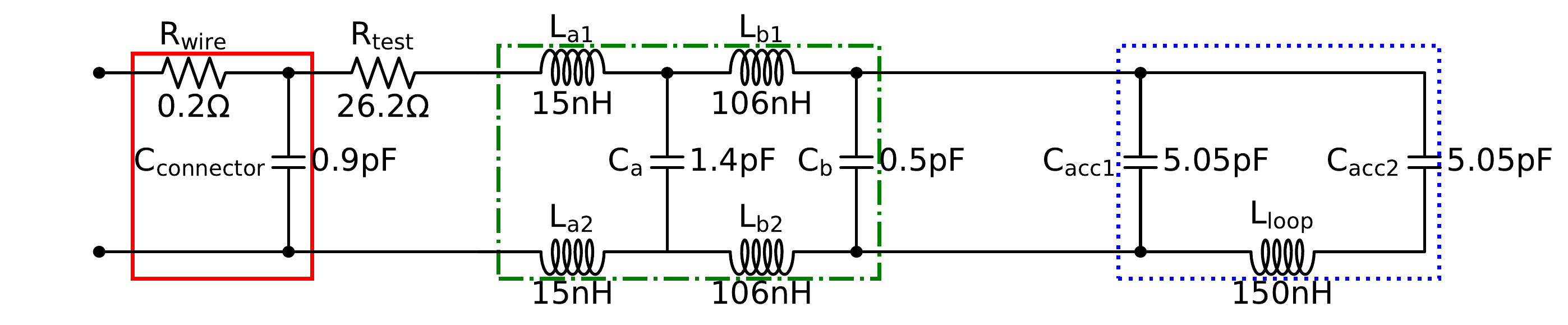}
\caption{The equivalent circuit used to model a RF unit including the cables. The two marked areas
on the left represent wires and connectors for the two RF-acceleration gaps modeled as two capacitors 
(right box) and an inductive cable. Several of these circuits in parallel were used to model 
scaling behaviour when driven by the RF source. The capacitor $R_{test}$ was only added during
some measurements to get better results from a vector network analyzer.}
\label{fig:wafer-schematic}
\end{figure}

The RF units were modeled as a series of cables with inductance and capacitance. The acceleration
gaps were modeled as purely capacitive and since the wafers are connected with a short wire loop, a
inductance was added between them. The complete circuit is shown in Figure~\ref{fig:wafer-schematic}.
The circuit model includes a capacitance to represent the BNC connector, a cable resistance (the cable
resistance was placed to the left of the capacitor due to requirements of the simulations software),
a test resistor (see below), a network of capacitors and inductors to represent the cable from the
RF source to the wafers, and the capacitance of the acceleration gap including an inductance that
represents the ground connection between the wafers.

To fit the parameters of the models, a RF unit was analyzed using a vector network analyzer (VNA),
Keysight Technology N9923A, in the frequency range of \SI{5}{\mega\hertz}-\SI{30}{\mega\hertz}
and also in the range \SI{5}{\mega\hertz}-\SI{150}{\mega\hertz}. Furthermore, the capacitance was
measured directly. The VNA measurement ($S_{11}$ mode) showed a resonance behaviour around \SI{90}{\mega\hertz}
which allowed us to calculate the inductance using the measured capacitance. Furthermore, since
we know the source impedance of the VNA (\SI{50}{\ohm}), we can calculate the load impedance at,
for example, \SI{15}{\mega\hertz} and from there estimate the resistance. However, due to the small
resistance of the system, it is difficult to measure this value directly. We therefore added
a known resistor, $R_{test}$, to the system, which made it easier to estimate the total resistance
using the VNA. The estimated value is in agreement with direct measurements and with calculated 
values when taking the skin effect into account.

Simulating the modelled circuits using LTspice~\cite{LTspice} and Matlab RF toolbox gave 
good agreement in the frequency range of operation (around \SI{15}{\mega\hertz}) with an
error between measurement and simulation smaller than $1\%$.

\begin{figure}[htbp]
\centering
\includegraphics[width=\linewidth]{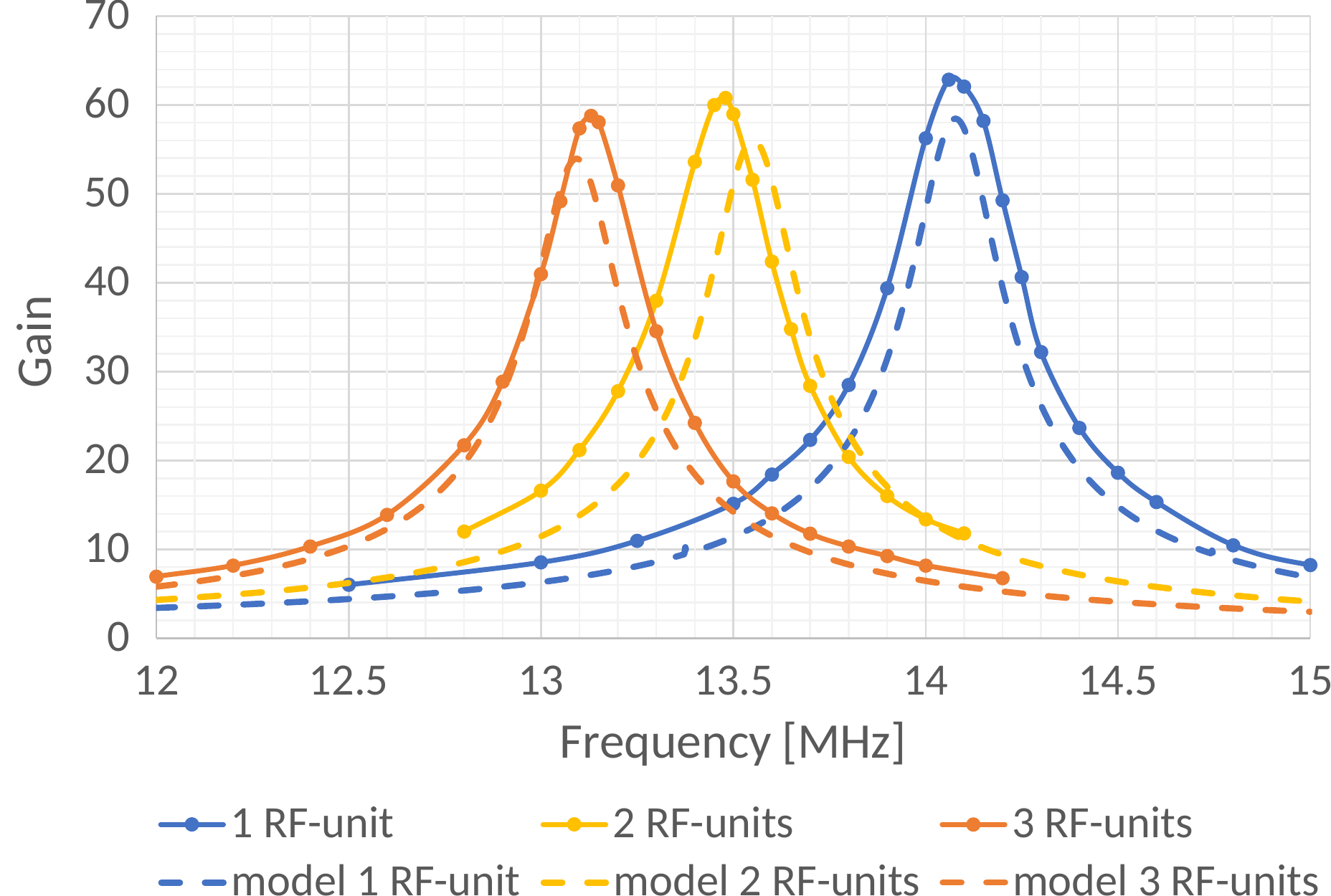}
\caption{Frequency response of the modelled and measured circuits using one to three RF units.}
\label{fig:model}
\end{figure}

Furthermore, the RF source allowed measurement of the output voltage using a voltage 
monitor output and we were also able to measure the voltage directly using a high
voltage probe (Tektronix P6015A). Figure~\ref{fig:model} compares the simulated and measured results when one to three RF units are used. The figure also shows that there is a slight shift in resonance frequency. This is due to the fact that we did not change the capacitor $C_2$ to optimally balance the load impedance of the RF source. This can also explain the slight decrease in gain between the different setups.

\section{Beam Experiment}
The beam energy was measured using the same setup and method as described previously\cite{Persaud:RSI-2016}. 
Ions were produced using a filament-driven ion 
source operated with argon at a pressure of \SI{1}{\pascal} (\SI{7.5}{\milli\torr}). The source was
floated at a high voltage of \SI{6}{\kilo\volt} and ions were extracted from a $3\times3$~array of extraction 
apertures using an extraction voltage of \SI{400}{\volt}. The ions are then accelerated to 
ground potential where they enter the RF units. Here, the ions are either accelerated or decelerated 
depending on their arrival time with respect to the phase of the RF voltage in the acceleration gaps. 
In our experiments we utilized a long pulse (\SI{300}{\micro\second}) compared to the 
RF frequency whereas in a final implementation of the accelerator one would bunch the beam and 
only inject beam packages at the correct phase. We then select ions above a certain beam energy using a 
deceleration grid and measure the beam current in a Faraday cup. Due to the almost continuous beam injection,
we expect a distribution of beam energies that we can model using a 1D simulation code. The length of the 
drift sections were chosen for optimal transport of the accelerated ions. We can therefore either fit the
beam energy using the simulation code or read of the maximal beam energy gain and divide this by the 
number of acceleration gaps to get the applied RF voltage. Experiments were carried out using one 
to three RF units.

\begin{figure}[htbp]
\centering
\includegraphics[width=\linewidth]{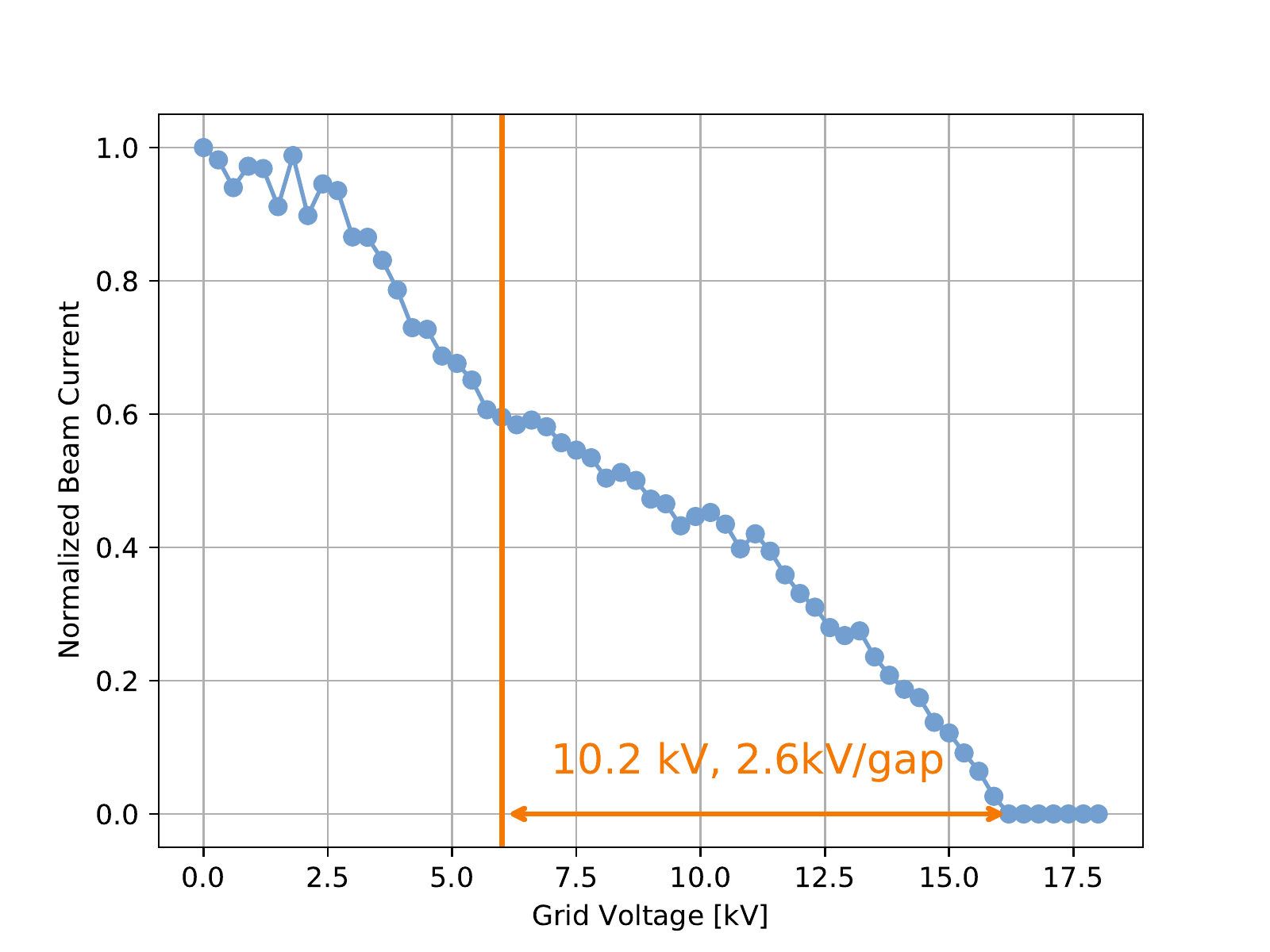}
\caption{Measured result of a grid voltage scan for a system using two RF units driving
the RF source with \SI{65}{\volt}. The measurement shows that \SI{2.6}{\kilo\volt} per
acceleration gap was achieved. The vertical line shows the beam energy before extraction (\SI{6}{\keV}).}
\label{fig:scan}
\end{figure}

Figure~\ref{fig:scan} shows the result of such a voltage scan on the retarding grid for a system using
two RF units (four acceleration gaps). The initial beam energy was \SI{6}{\keV} and one can clearly 
see that due to the continuous injection we measure all beam energies from 0-\SI{16}{\keV}. The
maximum energy gain was \SI{10.2}{\keV} equivalent to an applied voltage of \SI{2.56}{\keV} per
acceleration gap. The signal agrees well with our expectation from a 1d model \cite{Persaud:RSI-2016}.

\begin{figure}[htbp]
\centering
\includegraphics[width=\linewidth]{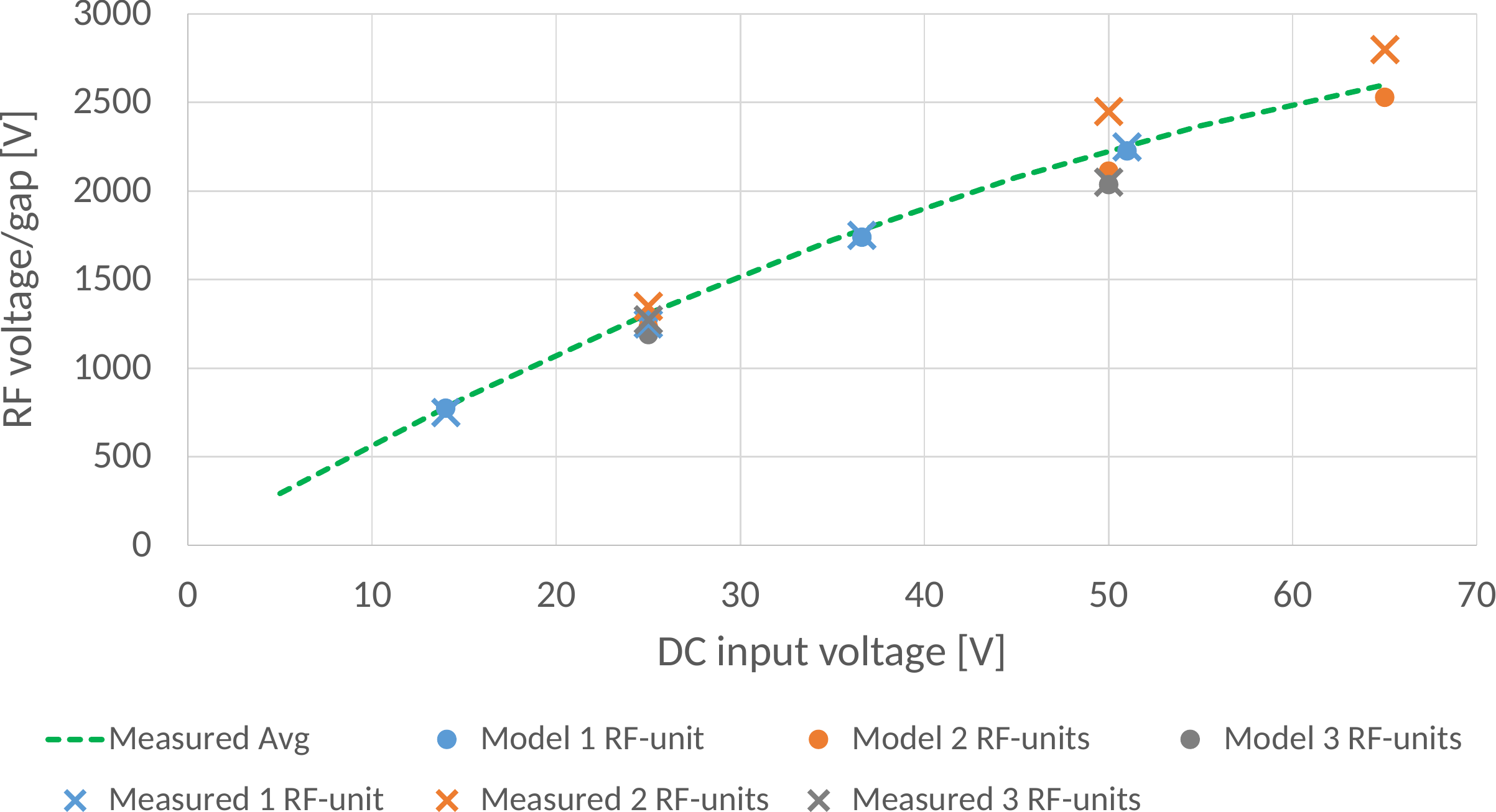}
\caption{Results from beam measurements showing the RF-output voltage per acceleration 
gap compared to the results from the expected result from the circuit model.}
\label{fig:result}
\end{figure}

\section{Outlook}
Using the circuit model we estimate that with the present RF source we will be able to drive 
nine RF-units in parallel by placing the RF source inside the vacuum system nearby the 
acceleration gaps. Simulations show that to achieve this one has to take care to minimize 
the capacitance in the cables connecting the RF units to the RF source. This is due to the 
fact that the capacitance of the cables for nine RF units will be a considerable part of 
the load that we need to drive. To minimize phase shift and cable length, we assume that
three commercial coax cables will be used in parallel from the RF source
which will then split into individual cables to each board. Our simulation was 
based on coax cables with an impedance of \SI{125}{\ohm} and a low capacitance. 

Furthermore, several possible optimization opportunities in our current design of the
RF units became clear. First of all, we should be able to further reduce the capacitance
of the acceleration gaps by reducing the amount of copper used and secondly higher-rated high-voltage connectors should be used to withstand the higher anticipated voltages per gap.
Custom RF cables could further lower the total load capacitance.

\section{conclusion}
Voltages of \SI{2.6}{\kilo\volt} per acceleration gap were achieved using a new, compact, near-board
RF driver. We believe that optimizing the RF driver specifically for this application will allow us
to double or triple the achieved voltage in the near future, and RF acceleration voltages of about
\SI{10}{\kV}/gap might be achievable in the future.
Furthermore, a model was developed and compared to measurements that allowed us to model scaling 
behaviour of the system to higher numbers of RF units. The simulations indicated that up to nine
RF units using the current setup can be driven by a single RF source in parallel. By further
optimizing the system we believe that this number can also be increased.

The demonstration of RF acceleration voltages of several kV/gap and the near-board RF driver is an 
important step in the development of a compact and low cost multi-beam accelerator technology 
that can be scaled to very high beam currents.  We estimate that we can accelerate ions in an 
array of at least $15\times15$ beams with a 4" wafer platform for a total current \SI{\sim 5}{\mA}
(from an ion source with current density of \SI{10}{\milli\ampere\per\square\cm}). With 
optimization of the RF drivers, acceleration gradients of over about \SI{1}{\mega\volt\per\meter}
can be implemented, resulting in a compact, high power accelerator for applications in 
materials processing.    

\section{acknowledgements}
We thank Michael Current for his encouragement and advice on the evolving field of high-energy ion implantation.

\bibliographystyle{IEEEtran}
\bibliography{IEEEabrv,paper}

\begin{thebibliography}{1}
\providecommand{\url}[1]{#1}
\csname url@samestyle\endcsname
\providecommand{\newblock}{\relax}
\providecommand{\bibinfo}[2]{#2}
\providecommand{\BIBentrySTDinterwordspacing}{\spaceskip=0pt\relax}
\providecommand{\BIBentryALTinterwordstretchfactor}{4}
\providecommand{\BIBentryALTinterwordspacing}{\spaceskip=\fontdimen2\font plus
\BIBentryALTinterwordstretchfactor\fontdimen3\font minus
  \fontdimen4\font\relax}
\providecommand{\BIBforeignlanguage}[2]{{%
\expandafter\ifx\csname l@#1\endcsname\relax
\typeout{** WARNING: IEEEtran.bst: No hyphenation pattern has been}%
\typeout{** loaded for the language `#1'. Using the pattern for}%
\typeout{** the default language instead.}%
\else
\language=\csname l@#1\endcsname
\fi
#2}}
\providecommand{\BIBdecl}{\relax}
\BIBdecl

\bibitem{Hamm12}
R.~W. Hamm and M.~E. Hamm, Eds., \emph{Industrial Accelerators and Their
  Applications}.\hskip 1em plus 0.5em minus 0.4em\relax World Scientific, 2012.

\bibitem{Brown2004}
I.~Brown, Ed., \emph{The physics and technology of ion sources}.\hskip 1em plus
  0.5em minus 0.4em\relax Wiley, 2004.

\bibitem{Maschke_1979b}
\BIBentryALTinterwordspacing
A.~Maschke, ``Meqalac: a new approach to low beta acceleration,''
  \emph{Technical report BNL-51029}, June 1979. [Online]. Available:
  \url{https://www.osti.gov/scitech/servlets/purl/5914442}
\BIBentrySTDinterwordspacing

\bibitem{URBANUS1989508}
\BIBentryALTinterwordspacing
W.~Urbanus, R.~Wojke, J.~Bannenberg, H.~Klein, A.~Schempp, R.~Thomae, T.~Weis,
  and P.~V. Amersfoort, ``Meqalac: A 1-mev multichannel rf-accelerator for
  light ions,'' \emph{Nuclear Instruments and Methods in Physics Research
  Section B: Beam Interactions with Materials and Atoms}, vol. 37-38, pp. 508
  -- 511, 1989. [Online]. Available:
  \url{http://www.sciencedirect.com/science/article/pii/0168583X89902346}
\BIBentrySTDinterwordspacing

\bibitem{Persaud:RSI-2016}
\BIBentryALTinterwordspacing
A.~Persaud, Q.~Ji, E.~Feinberg, P.~A. Seidl, W.~L. Waldron, T.~Schenkel,
  A.~Lal, K.~B. Vinayakumar, S.~Ardanuc, and D.~A. Hammer, ``A compact linear
  accelerator based on a scalable microelectromechanical-system rf-structure,''
  \emph{Review of Scientific Instruments}, vol.~88, p. 063304, 2017. [Online].
  Available: \url{http://aip.scitation.org/doi/10.1063/1.4984969}
\BIBentrySTDinterwordspacing

\bibitem{Seidl2018}
\BIBentryALTinterwordspacing
P.~A. Seidl, A.~Persaud, W.~Ghiorso, Q.~Ji, W.~L. Waldron, A.~Lal, K.~B.
  Vinayakumar, and T.~Schenkel, ``Source-to-accelerator quadrupole matching
  section for a compact linear accelerator,'' \emph{Rev. Sci. Instrum.},
  vol.~89, no.~5, p. 053302, May 2018. [Online]. Available:
  \url{https://doi.org/10.1063/1.5023415}
\BIBentrySTDinterwordspacing

\bibitem{Vinaya2018}
K.~Vinayakumar, A.~Persaud, Q.~Ji, P.~Seidl, T.~Schenkel, and A.~Lal,
  ``Waferscale electrostatic quadrupole array for multiple ion beam
  manipulation,'' in \emph{IEEE proceedings of MEMS 2018}, 2018.

\bibitem{Airity}
{Airity Technologies, Inc.}, https://www.airitytech.com/, 2018.

\bibitem{LTspice}
{Analog Devices},
  http://www.analog.com/en/design-center/design-tools-and-calculators/ltspice-simulator.html,
  2018, [Online; accessed May 2018].

\end{thebibliography}

\end{document}